\documentstyle[mprocl,epsfig]{article}
\def\be{\begin{equation}}
\def\ee{\end{equation}}
\def\bea{\begin{eqnarray}}
\def\eea{\end{eqnarray}}
\begin{document}

\title{CORRELATIONS WEAK AND STRONG: DIVERS GUISES OF
THE TWO-DIMENSIONAL ELECTRON GAS}

\author{A.H. MACDONALD}

\address{Physics Department, Indiana University,
Bloomington, IN 40475, USA\\E-mail macdonal@indiana.edu}


\maketitle\abstracts{
The three-dimensional electron-gas model has been a major focus 
for many-body theory applied to the electronic properties of metals
and semiconductors.  Because the model neglects band 
effects, whereas electronic systems are generally more strongly correlated 
in narrow band systems, it is most
widely used to describe the qualitative physics of weakly correlated metals
with unambiguous Fermi liquid properties.  
The model is more interesting in two space dimensions because it 
provides a quantitative description of electrons in quantum wells
and because these can form strongly correlated many-particle states.
We illustrate the range of possible many-particle behaviors 
by discussing the way correlations are manifested in 
2D tunneling spectroscopy experiments. 
}

\section{Introduction} 

In metals and semiconductors electrons interact among themselves
and with nuclei which are localized, in ordered solids, near lattice 
sites.  Even after ignoring the translational degrees of 
freedom of the nuclei, we are left with a host of different
many-body problems for interacting electron systems.
In elemental solids, we have $\sim 100$ different 
interacting electron problems specified, for example, by the 
atomic numbers and lattice constants of the periodic table 
on the inside front cover of the text by Ashcroft and
Mermin.\cite{ashcroftmermin}  Problems of current interest 
in many-electron physics are more likely to be found in 
complicated compounds such as the quaternary cuprate 
superconductors, or in artificially fabricated systems such as  
metallic magnetic multilayers or semiconductor quantum dots.
The set of possible many-electron problems is endless
and, fortunately for this scientific subfield, the 
history of condensed matter physics teaches 
that discoveries of unanticipated properties in newly 
synthesized materials will also be endless. 

Given the large number of many-electron problems,
the hopeless difficulty of each, and 
the difference between the largest energy
scales of these problems and the energy scales
of the excitations that determine physical properties
at room temperature and below, a common strategy of 
many-body theories is to work with models that are intended
to faithfully describe essential aspects of the observed behavior 
of classes of materials without introducing incidental complications.
Perhaps the most commonly studied models are the 
Hubbard model, which emphasizes the physics of 
strong repulsion between electrons near the same 
lattice site, and the electron gas model 
which mistreats atomic-like correlations
at short length scales and captures instead the 
physics of screening and long-length scale correlations
in metals.  All electron {\it ab-initio} calculations 
are generally limited to the consideration of 
physical properties and systems for which
self-consistent-field approximations, still 
present in practice even in modern
density-functional\cite{densfunc} based
electronic-structure calculations, are adequate. 
This situation frequently creates a difficulty in judging 
whether discrepancies between many-body theories and 
experiment are due to inadequacies of a model, or due
to inadequacies of the approximations used to estimate
the properties of a model.  The case of electrons
in semiconductor quantum wells is unusual in this 
respect.  At energies below the subband splitting of the 
quantum well, these systems are extremely accurately 
modeled by a two-dimensional (2D) version of the
electron-gas model.  In this paper we discuss some recent experimental
and theoretical work on the many-body physics of the 
2D electron gas.  We limit our attention
to the one-particle Greens function, which can
be probed experimentally in these systems by 2D-2D tunneling
spectroscopy.\cite{2d2dtunexpt,ashoori,2d2dtun} 

The paper is organized as follows.  In section II we 
discuss the case of a 2D electron system in
a strong magnetic field with a partially filled 
Landau level.  In this limit, the electronic ground
state is not a Fermi liquid and the one-particle Greens function
exhibits strongly non-perturbative behavior in which a large gap surrounding
the Fermi energy is present in the 
tunneling density-of-states.
This behavior is characteristic of strongly correlated
electronic states.  In section III we discuss the 
case of zero magnetic field, where experimental 
results confirm the Fermi-liquid nature of the 
electronic state, and tunneling experiments 
imply quasiparticle lifetimes in semi-quantitative
agreement with simple random-phase-approximation
estimates.  Section IV offers some
speculations on changes that should be expected
in the zero-field tunneling density of states as
the electron density is lowered and the electronic system
becomes more strongly correlated.  We conclude in Section V
with a brief sumamry. 

\section{One-particle Greens Function at Strong Magnetic Fields} 

For a non-interacting electron system, the one-particle
Greens function $G_i(\epsilon)$ has a single pole 
at $\epsilon = \epsilon_i$, where $\epsilon_i$ is an
eigenvalue of the one-body Hamiltonian.
Since many observables can be expressed entirely in
terms of the one-particle Greens function,\cite{mbtexts}  
the way in which it is altered by interactions is a central topic 
of many-body theory. In 2D systems tunneling experiments\cite{2d2dtun}  
have proved particularly valuable for experimental
studies of the one-particle Greens function.  In a
tunneling-Hamiltonian formalism,\cite{wolf} the tunneling current 
between weakly linked electronic systems is given at zero temperature by 
\begin{equation}
I = \frac{2 \pi e}{\hbar} \sum_{i_L,i_R}
 |t(i_L,i_R) |^2 \int_{\mu}^{\mu + eV}  d \epsilon 
A^{+}_{i_L}(\epsilon) A^{-}_{i_R}(\epsilon - e V).
\label{tuncond}
\end{equation}
where $\mu$ is the Fermi energy,
$L$ and $R$ labels are used to distinguish 2D 
systems on opposite sides of the tunneling barrier,
$A_{i}^{+}(\epsilon)$ and $A_{i}^{-}(\epsilon)$ are the 
particle and hole contributions to the spectral weight function 
for single-particle label $i$ and $V$ is the bias voltage.
The spectral weights are related to the exact one-particle 
Greens function as described below.  When interactions are 
neglected $A_{i}^{+}(\epsilon) = \delta (\epsilon - \epsilon_i)$ 
if state $i$ is empty, $A_{i}^{-}(\epsilon) = \delta(\epsilon - \epsilon_i)$
if state $i$ is occupied and the spectral weights are otherwise zero. 
In strongly correlated systems, the spectral function can be 
qualitatively altered.  

In the presence of a perpendicular magnetic field, the 
single-particle kinetic-energy spectrum consists of 
Landau levels\cite{leshouches} with energy $\epsilon_n = 
\hbar \omega_c (n + 1/2)$ and macroscopic degeneracy 
$N_{\phi} = A B / \Phi_0$.  Here $\omega_c = e B/ m c$ is
the cyclotron frequency, $A$ is the system area, $B$ is
the magnetic field strength, and $\Phi_0 = hc/e$ is the 
magnetic flux quantum.  The one-particle Greens function is defined by 
\begin{equation}
G_n(\epsilon) = \int_{- \infty}^{\infty}  \frac{dt}{\hbar} \exp (i \epsilon t / \hbar ) G_n(t)
= -i \int_{-\infty}^{\infty} \frac{dt}{\hbar} \exp (i \epsilon t / \hbar )   
 \langle \Psi_0 | T [ c_{n,m}(t) c^{\dagger}_{n,m}(0) ] | \Psi_0 \rangle
\label{gfsf}
\end{equation} 
where $c_{n,m}$ and $c_{n,m}^{\dagger}$ are respectively fermion 
annihilation and creation operators for one of the degenerate 
states within the $n$'th Landau level, and $|\Psi_0\rangle$ is the 
ground state of the interacting electron system.\cite{caveatgf}  
We restrict our attention here to $n=0$ and drop the 
Landau level index on $G_{n}(\epsilon)$.  The Greens function can then
be written in the form
\begin{equation}
G(\epsilon) = \int d \epsilon ' \big[ 
\frac{A^{+}(\epsilon ')}{\epsilon - \epsilon ' + i \eta} 
+ \frac{A^{-}(\epsilon ')}{\epsilon -\epsilon ' - i \eta} \big] 
\label{gfsrep}
\end{equation}
where the particle and hole spectral functions $A^+$ and $A^-$ have
the following formal expression in terms of exact eigenstates of the 
Hamiltonian: 
\begin{eqnarray}
A^+(\epsilon)&=&\sum_{\alpha}|\langle \Psi_{\alpha} (N+1)
|c^{\dagger}_m|\Psi_0(N) \rangle|^2 \nonumber \\
&&  \quad \times \delta(\epsilon-[E_{\alpha}(N+1) - E_0 (N)] ) \nonumber\\
\noalign{\vskip0.2cm}
A^-(\epsilon)&=&\sum_{\alpha}|\langle | \Psi_{\alpha} (N-1) |c_m|\Psi_0(N)
\rangle  |^2 \nonumber \\
&&  \quad \times \delta(\epsilon-[E_0(N) - E_{\alpha}(N-1)])
\label{eq:tdosdef}
\end{eqnarray}
\begin{figure}[ht]
\epsfysize=3.5in
\centerline{\epsfbox{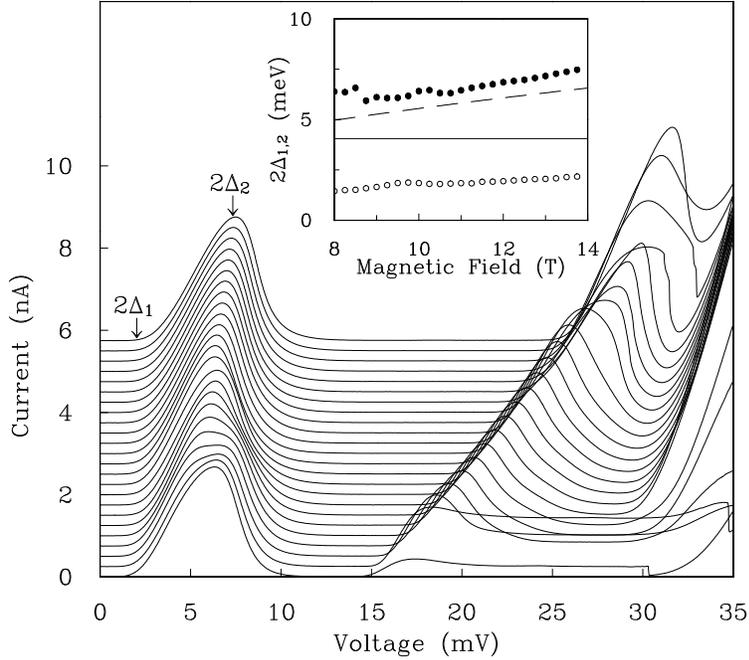}}
\vspace{6pt}
\caption[]{Low temperature tunneling I-V characteristics for bulk 2D to 2D
tunneling in the fractional Hall regime.  The traces are
at magnetic fields separated by $0.25$ Tesla and for
this sample cover a range of filling factors from
$\nu = 0.48$ to $\nu = 0. 83$.  The higher bias
potential peak is due to Landau level mixing while the lower
bias potential peak is due to tunneling within the lowest
Landau level.  The tunneling current is the convolution of
particle and hole spectral weight functions, at energies 
separated by $e V$.  This experiment shows that the tunneling current
is strongly suppressed near zero bias and hence that the
spectral weight functions are suppressed at energies near the
Fermi energy over a broad range of filling factors.  Features
in the tunneling data associated with particular
incompressible states, for example the one that occurs at $\nu = 2/3$, are
weak. The inset shows the onset and peak values of the intra
Landau level tunneling currents.
(After Eisenstein \textit{et al.} in Ref.11.)
}
\label{fig1}
\end{figure} 

When interactions are neglected, states in the lowest Landau level
are degenerate and are occupied with probability $\nu$ where $\nu
= N/N_{\phi}$ and $N$ is the number of electrons in the many-particle system.
It follows that $A_{+}(\epsilon) = (1 - \nu) 
\delta (\epsilon)$ and $A_{-}(\epsilon)=\nu \delta(\epsilon)$.  
When these spectral functions are inserted in Eq.(~\ref{tuncond})
we find that $I \propto \nu (1 - \nu) \delta (eV)$.  If interactions 
were unimportant, a sharp peak in the tunneling current would occur
near zero bias.  The experimental results\cite{tundossfe} of Eisenstein
{\it et al.} shown in Fig.[1] evidence quite 
the opposite behavior; the tunneling current is immeasurably small
at zero bias and is peaked instead 
near a bias voltage of $\sim 5 {\rm mV}$.  Since $I(V)$ depends on the 
product of electron and hole spectral functions at energies within
$eV$ of the Fermi energy, it follows that both are extremely small
close to the Fermi energy.  Similar conclusions can be drawn on the 
basis of 3D-2D tunneling experiments\cite{ashoori,tundossfa} 
by Ashoori and collaborators.
This experimental
result can be understood in qualitative terms 
as a consequence of strong correlations
in the ground state of the electron system.  When an additional 
electron is added to the ground state 
of an $N$ electron system, it is 
not strongly correlated with the electrons already present.
The resulting $N+1$ particle state will have a small overlap with the
ground state or any low energy states of the strongly correlated $N+1$ electron
system.  $A_{+}(\epsilon)$ should therefore be peaked at energies well above the 
energy difference between the ground states of the $N+1$ and 
$N$ particle systems, the chemical potential $\mu$.

Despite the achievement of a number of valuable insights,\cite{sftheory}
we still do not have a completely satisfactory theory for 
the spectral weight function in the strong magnetic field limit.
Nevertheless, the connection between the energy at which the tunneling 
current is peaked and correlations in the ground state of 
the electron system suggested by the preceding words 
can be made more precise.\cite{haussmann,renn}  The following sum rules 
for the zeroth and first moments of 
$A(\epsilon) \equiv A_{-}(\epsilon) + A_{+}(\epsilon) $ \,
follow from Eqs.(~\ref{eq:tdosdef}): 
\begin{equation} 
\int d \epsilon A(\epsilon) =
\langle \Psi_0 | c_m^{\dagger} c_m + c_m c_m^{\dagger} | \Psi_0 \rangle = 1 
\label{zeroth}
\end{equation}
and
\begin{eqnarray} 
\int d \epsilon \epsilon A(\epsilon) 
&=&\langle \Psi_0 | [H,c_m^{\dagger}] c_m +
c_m [H, c_m^{\dagger}] | \Psi_0 \rangle \nonumber \\ 
&=& \epsilon_{HF}.
\label{first}
\end{eqnarray} 
In Eq.(~\ref{first}),
\begin{equation} 
\epsilon_{HF} = \nu \sum_{m'}[ \langle m,m' | V | m, m' \rangle
- \langle m',m | V | m, m' \rangle ] 
\label{hfock}
\end{equation}
is the single-particle Hartree-Fock energy for this system.
The last identity
in Eq.(~\ref{first}) follows from translational invariance 
which requires that $\langle \Psi_0 | c_m^{\dagger} c_m | \Psi_0 \rangle
= \nu $ for all $m$.  These sum rules should be compared with those 
for $A_{-}$:
\begin{equation}
\int d \epsilon A_{-}(\epsilon) = \langle \Psi_0 |
c_m^{\dagger} c_m | \Psi_0 \rangle =  \nu 
\label{mzeroth}
\end{equation}
and 
\begin{equation} 
\int d \epsilon \epsilon A_{-}(\epsilon) 
=\langle \Psi_0 | [H,c_m^{\dagger}] c_m | \Psi_0 \rangle  = 2 \epsilon_0(\nu). 
\label{mfirst}
\end{equation}
where $\epsilon_0(\nu) = E_0/N_{\phi}$ is the many-particle ground state energy
per lowest-Landau-level orbital.  The last identity is an adaptation, to  
the case of Hamiltonians that include only an interaction term,
of the expression for the ground state energy in terms of the
one-particle Greens function, derived using an equation of motion
approach in many-body theory texts.\cite{mbtexts}  Its application here rests  
on the observation that
$\langle \Psi_0 | [H,c_m^{\dagger}] c_m | \Psi_0 \rangle$ is independent of $m$.  
The integrated weight of the hole part of the spectral weight function 
is $\nu$ and its mean energy is therefore 
$ \langle \epsilon_{-} \rangle  = 2 \epsilon_0(\nu)/\nu$.
For $A(\epsilon)$, the integrated weight is $1$ and the 
mean energy is $\epsilon_{HF}$.   It follows that the mean energy in the particle
part of the spectral weight function is 
\begin{equation}
\langle \epsilon_{+} \rangle  = \langle \epsilon_{-} \rangle +
\frac{ \nu \epsilon_{HF} - 2 \epsilon_0(\nu)}{\nu (1 - \nu)} 
\label{gap}
\end{equation} 
The second term on the right hand side of this equation is a sum rule
estimate of the gap separating peaks in the hole and particle
portions of the spectral weight, $\Delta_{sr}$ . 
$\Delta_{sr}$ will give an accurate value for the peak position
whenever the peak is sharp and its position well defined;  
a property established for the present case by experiment.
The dependence of $\Delta_{sr}$ on $\nu$ is illustrated in Fig.[2]; when
corrected for finite quantum well width effects these results agree
well with experiment.
\begin{figure}[ht]
\epsfxsize=3.5in
\centerline{\epsfbox{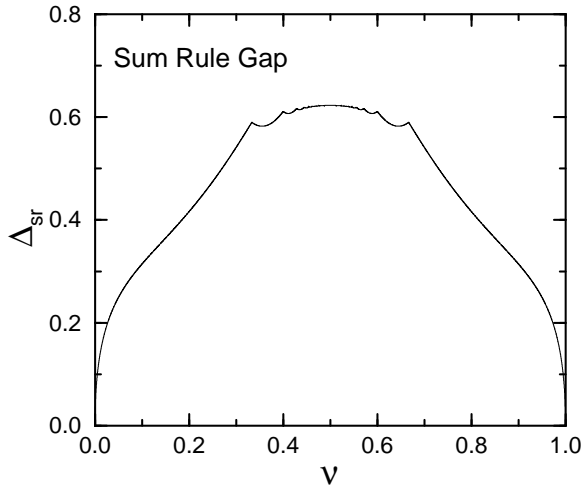}}
\vspace{6pt}
\caption[]{Sum rule estimate for the gap between 
particle and hole contributions to the spectral weight 
function of a two-dimensional electron gas at strong
magnetic fields as a function of Landau level 
filling factor $\nu$.  The tunneling conductance is peaked when
$eV = \Delta_{sr}$.  These estimates are for the case of 
an ideal two-dimensional electron gas and need to be
reduced to account for finite thickness effects when
comparing with experimental systems.  The gap 
is expressed in units of $e^2/\epsilon \ell$ where 
$\ell = (\hbar c/eB)^{1/2}$ is the magnetic length.
After Mori {\it et al.} in Ref. 14.}
\label{fig2}
\end{figure}

Since in the present case there is no one-body term in the Hamiltonian, 
the ground state energy per orbital in the Hartree-Fock approximation 
is $\nu \epsilon_{HF}/2$.  The gap is proportional to the amount by
which the ground-state energy lies below the Hartree-Fock approximation
ground-state energy, {\it i.e.}, the gap is proportional to the correlation
energy.  We see that in the strong-magnetic-field case where the kinetic
energy can be completely removed from the problem, the intuitive 
notion that there should be a gap in the electronic spectral function
related to ground-state correlations has surprising exactness.
Now we turn to the zero magnetic field limit, where interactions
are in competition with the kinetic energy.  

\section{2D Tunneling Spectroscopy of the Fermi Liquid State}

At zero magnetic field, we can label states
in the 2D electron system by two-dimensional wavevectors.  A
unique feature of 2D-2D tunneling between quantum wells in
epitaxial layered semiconductors, is the near perfect translational invariance
perpendicular to the tunneling direction which selects
momentum-conserving tunneling processes.  Accounting for 
spin-degeneracy at zero field, the tunneling current is
\begin{equation}
I = \frac{4 \pi e |t|^2 }{\hbar} \sum_{\vec k}
  \int_{\mu}^{\mu + eV}  d \epsilon 
A^{+}_{\vec k}(\epsilon) A^{-}_{\vec k}(\epsilon - e V).
\label{2dtuncond}
\end{equation}
For non-interacting 2D electrons, the spectral-weight function
had a delta-function peak in the hole contribution for $|\vec k| < k_F$ and 
a delta-function peak in the particle contribution for $|\vec k| > k_F$
where $k_F$ is the Fermi wavevector.
The tunneling current consequently has a delta-function peak at zero
bias.  At zero magnetic field, it is generally expected that the
interacting 2D electron system should be in a
Fermi-liquid state in which a large fraction of the spectral weight lies in 
a sharp Lorentzian quasiparticle peak centered at the 
quasiparticle energy:
\begin{equation}
A_{\vec k}(\epsilon) = \frac{z_{\vec k}}{\pi}\frac{\Gamma_{\vec k}}{(\epsilon - E_{\vec k})^2 + \Gamma_{\vec k}^2} + A^{inc}_{\vec k}(\epsilon) 
\label{flspec}
\end{equation}
Here $\Gamma_{\vec k} = \hbar / 2 \tau_{\vec k}$ where $\tau_{\vec k}$ is the 
quasiparticle lifetime, $E_{\vec k}$ is the 
quasiparticle energy, $z_{\vec k} < 1 $ is the quasiparticle normalization
factor, and $A^{inc}_{\vec k}$ is the part of the spectral function
non included in the quasiparticle peak.  In a Fermi liquid, the quasiparticle 
lifetime diverges in the limit of zero temperature as $k$ approaches 
$k_F$.

If we ignore the renormalization of the quasiparticle energy due 
to interactions and the decrease of the quasiparticle lifetime 
on moving away from the Fermi surface, the contribution to
the integrals in Eq.(~\ref{2dtuncond}) from the quasiparticle portion
of the spectral function can be evaluated analytically with the result
\begin{equation}
G^{qp}(V) = \frac{I(V)}{V} = \frac{z_{k_F}^2 e^2 |t|^2 A \nu_0}{\hbar}
\frac{ 2 \hbar / \tau}{ (e V)^2 + ( \hbar/ \tau)^2} 
\label{tuncondfl}
\end{equation} 
where $\nu_0$ is the 2D Fermi-gas density of states.  A finite
quasiparticle lifetime turns the $\delta$ function peak in
the tunneling conductance into a Lorentzian.  This is exactly what is 
seen in experiment,\cite{murphybzero} providing a very direct 
confirmation of the Fermi-liquid nature of the 2D electron-gas
state at zero field.  The zero-bias {\it peak} in the tunneling 
conductance at zero field 
stands in stark contrast with the extremely small conductance at 
small bias found at strong fields and discussed in the previous section.  
The 2D tunneling-spectroscopy experiment is a sensitive indicator 
of the nature of the many electron state.
This simple result for the tunneling-conductance-spectroscopy line shape 
applies when all quasiparticle lifetimes are approximated by their
value at $|\vec k| = k_F$.  Detailed calculations\cite{jungwirth,zheng}
show that the true line shape is not Lorentzian, but that 
the voltage at which $G^{qp}$ is reduced to half of 
its $V=0$ value is increased by only $20\%$ when the $\vec k$ 
dependence of the quasiparticle lifetime is taken into account.
Thus it is possible to read off the quasiparticle lifetime at the Fermi energy by
looking at the width of the zero-bias tunneling conductance peak.
As illustrated in Fig.[3], the measured lifetimes are in
reasonable accord with random-phase-approximation theoretical 
estimates.   Theoretical and experimental lifetimes can
be brought into closer agreement by incorporating proposals  
for improving on the random-phase approximation from the 
electron-gas many-body-theory literature.  The improved 
agreement supports the 
efficacy of exchange-correlation local-field corrections 
to the random phase approximation.\cite{locfield}
\begin{figure}[ht]
\epsfxsize=3.5in
\centerline{\epsfbox{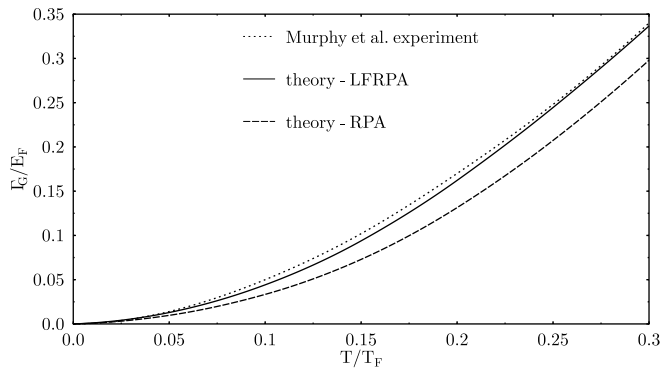}}
\vspace{6pt}
\caption[]{ Half-width at half-maximum for the zero-bias 2D-2D 
tunneling spectroscopy peak in the Fermi liquid state of a 2D
electron gas.  This width is inversely proportional to the quasiparticle
lifetime at the Fermi energy.  Theoretical results including
local-field corrections to the RPA (full
line), theoretical results in the RPA (dashed line), and experimental 
results (dotted line) reported by Murphy {\it et al.} in Ref. 17.
After Jungwirth {\it et al.} in Ref. 17.}
\label{fig3}
\end{figure}

The sharp peak in the tunneling conductance near zero bias 
is due entirely to quasiparticle peaks in the spectral weight  
functions.  As long as the 2D electron system is a Fermi liquid, 
this peak will be present.  However, as the electron system 
becomes more strongly correlated, the quasiparticle normalization
factor $z_{\vec k}$ will become smaller and the incoherent part
of the spectral function will become more dominant.  We now turn our 
attention to some speculations on what will happen in this regime.

\section{Strong Correlations at Zero Magnetic Field} 

A 2D electron system becomes increasingly correlated at low electron
densities.  The density is usually parameterized in terms of the 
electron gas parameter $r_s$ defined by the following equation:
\begin{equation}
n = N/A = (\pi r_s^2 a_0^2 )
\label{2drs}
\end{equation} 
where $a_0 = \hbar^2 \epsilon / m^* e^2$  is the host semiconductor 
Bohr radius, $m^*$ is the band-structure effective mass of the 
electrons, and $\epsilon$ is the host semiconductor  
dielectric constant.  When $r_s$ is large, the typical
interaction-energy scale is much larger than the typical kinetic-energy 
scale, and the electron system is strongly correlated.  For $r_s$ 
larger than\cite{wcrefs} about $30$ a phase transition between the 
Fermi-liquid electronic state and a Wigner-crystal state with broken
translational symmetry is expected to occur.  Existing tunneling 
experiments in 2D electron systems have been performed on samples
with $r_s < 2$, which have moderate correlations adequately described
by the random-phase approximation.  We expect that when experiments
are performed in samples with much lower density, the zero-bias
Fermi-liquid peak in the tunneling conductance will be 
embedded in a broader structure with a minimum near zero bias
similar to what has already been observed for strongly correlated
states in partially filled Landau levels.  This expectation follows
from a sum-rule analysis similar to that in Section II.

At zero magnetic field, the one-particle Greens function is 
diagonal in a representation of wavevectors and spins,
$\vec k$ and $\sigma$.  Although, unlike the strong field case, 
the Greens function is dependent on its single-particle state
labels, it still satisfies similar identities for the zeroth and 
first moments of the full spectral function:
\begin{equation}
\int_{- \infty}^{\infty} A_{\vec k, \sigma} (\epsilon) = 
\langle \Psi_0 | c_{\vec k,\sigma}^{\dagger} c_{\vec k, \sigma}  +
c_{\vec k \sigma} c_{\vec k \sigma} ^{\dagger} | \Psi_0 \rangle = 1 
\label{b0zeroth}
\end{equation}
and
\begin{eqnarray} 
\int d \epsilon \epsilon A_{\vec k,\sigma}(\epsilon)  &=&
\langle \Psi_0 | [H,c_{\vec k,\sigma}^{\dagger}] c_{\vec k, \sigma}  +
c_{\vec k \sigma} [H,c_{\vec k \sigma} ^{\dagger}] | \Psi_0 \rangle \nonumber \\
&=& \epsilon_{\vec k} + \frac{N}{A} V(\vec q = 0) 
- \frac{1}{A} \sum_{\vec k'} V(\vec k - \vec k') n_{\vec k',\sigma} \nonumber \\ 
&=& E_{\vec k}
\label{b0first}
\end{eqnarray} 
Note that in Eq.(~\ref{b0first}), $n_{\vec k,\sigma} = 
\langle \Psi_0 | c_{\vec k,\sigma}^{\dagger} c_{\vec k, \sigma} | \Psi_0 \rangle$ 
rather than the non-interacting Fermi gas occupation number.  If not for this
distinction, $E_{\vec k}$ would be the Hartree-Fock approximation
to the quasiparticle energy.  Below we compare the total energy of
the electron system with the quantity
\begin{equation}
\tilde E \equiv \frac{1}{2} \sum_{\vec k,\sigma} n_{\vec k,\sigma}
(\epsilon_{\vec k} + E_{\vec k}).
\label{etilde}
\end{equation}
If $n_{\vec k}$ were equal to its non-interacting value, 
$\tilde E$ would be the Hartree-Fock approximation to the 
ground-state energy.  Actually, $\tilde E$ will be larger than the 
Hartree-Fock ground-state energy, since the kinetic energy is increased
and the exchange energy reduced in magnitude 
by the partial occupation of states outside the Fermi sea.  
The difference between the exact ground-state energy $E$ and 
the approximate value obtained in a Hartree-Fock
approximation is generally referred to as the correlation energy
of an interacting electron system.  In the following, we will
appropriate this language in referring to the difference between $E$ and 
$\tilde E$.

We employ an exact identity\cite{mbtexts} that follows from the 
equation of motion for the one-particle Greens function
and relates the hole part of its spectral weight to the 
total energy: 
\begin{equation}
E = \frac{1}{2} \sum_{\vec k,\sigma} n_{\vec k}\big[ \epsilon_{\vec k}
+ \langle \epsilon_{\vec k}^{-} \rangle  \big] .
\label{totalenergy}
\end{equation} 
Here we have noted that $n_{\vec k}$ is the zeroth moment of
$ A_{\vec k}^{(-)}(\epsilon) $ and defined the average hole 
energy $\langle \epsilon_{\vec k}^{-} \rangle $ as the ratio of the first and 
zeroth moments.  Comparing this expression with the expression for
$\tilde E$, we find that the correlation energy can be written in the 
form:
\begin{equation}
E_{corr} \equiv E - \tilde E = \frac{-1}{2} \sum_{\vec k,\sigma} 
n_{\vec k} (1 - n_{\vec k}) (\langle \epsilon_{\vec k}^{+} \rangle -
\langle \epsilon_{\vec k}^{-} \rangle ) 
\label{correnergy}
\end{equation} 
where $\langle  \epsilon_{\vec k}^{+} \rangle $ is the mean energy in the particle
portion of the spectral weight function so that 
\begin{equation}
n_{\vec k} \langle  \epsilon_{\vec k}^{-} \rangle  + (1 - n_{\vec k})
\langle  \epsilon_{\vec k} ^{+} \rangle = E_{\vec k}. 
\label{eplusbar}
\end{equation} 
For a low-density 2D electron gas, correlations are strong.  The correlation
energy is negative and comparable in magnitude to the total 
energy per electron.  One way in which this can be consistent with 
Eq.(~\ref{correnergy}) is if, for $\sim N$ values of $\vec k$,
the spectral-weight function $A_{\vec k}(\epsilon)$
has a substantial portion of its weight distributed between
{\it non-quasiparticle} peaks located near
$\langle \epsilon_{\vec k}^{-} \rangle $ and 
$\langle \epsilon_{\vec k}^{+} \rangle $ and relatively little weight attached to 
its dispersive quasiparticle peak.  This is precisely what happens 
in the strong field limit where, in accord with the non-Fermi-liquid
character of the ground state, the quasiparticle peak is entirely absent.
We now consider two quite 
different examples, in which this scenario is realized.  
The generality of the behavior convinces us that it is 
ubiquitous in strongly correlated fermion systems.  

The one-band Hubbard model is a lattice model in which 
electrons interact only if they occupy the same lattice 
site.  The one-band version of this model has been
extremely widely studied\cite{elbio} in connection with the strong 
correlations that exist in the planar cuprate superconductors,
especially in the underdoped regime.  Electrons in this model become strongly
correlated when the band is near half-filling and 
when the on-site interaction $U$ is larger than 
the intersite hopping energy $t$.  The one-particle
Greens function is readily evaluated at half-filling 
in the narrow-band ($t \to 0$) limit.  The ground state has eigenenergy
$0$ and is  $2^{N}$ fold degenerate.  The zero energy states 
are those with one 
electron of either spin on each lattice site.  Averaging over
these states gives $n_{\vec k,\sigma} = 1/2$.
Removing an electron from any lattice site in any of the $2^{N}$ states
gives a state with zero energy so that 
for each wavevector in the Brillouin zone,
$A_{\vec k,\sigma}^{-}(\epsilon) = (1/2) \delta (\epsilon ) $ and 
$\langle \epsilon_{\vec k}^{-} \rangle = 0$.  Adding an electron on any lattice 
site produces a state with energy $U$ so that 
$A_{\vec k,\sigma}^{+}(\epsilon) = (1/2) \delta (\epsilon -U ) $ and
$\langle \epsilon_{\vec k}^{+} \rangle = U$.  These two peaks in the 
spectral function are referred to as the lower and upper Hubbard 
bands and they broaden when $t \ne 0$.  The mean-field single-particle 
energy, calculated from the mean occupation numbers but including 
exchange corrections which eliminate interactions between parallel
spins on the same site, is $E_{\vec k} = U/2$ and, taking note of  
the double-counting factor for this interaction energy, 
the corresponding total energy is $\tilde E = N U/4$.
In this case, the spectral weight is entirely comprised of 
non-dispersive delta-function hole and particle contributions.
This non-Fermi liquid state has no dispersive quasiparticle
pole which sharpens on crossing the Fermi energy. 
In precise agreement with Eq.(~\ref{correnergy}),  the two peaks 
are split by an amount proportional to the correlation energy
per particle.  For finite $t$, the Hubbard model is generally 
expected to have a Fermi liquid ground state away from the half-filled
band condition.  An important issue\cite{elbio} in the many-body theory
of the Hubbard model has been the question of whether the dispersive
quasiparticle peaks
at the Fermi energy are created by moving one of the Hubbard bands to the
Fermi energy or by drawing weight away from the Hubbard bands to 
quasiparticle bands at the Fermi energy.  Since the correlation
energy can be strongly dependent on band-filling, the sum rules 
discussed here appear to preclude the former possibility and argue
for quasiparticle bands which are energetically separate from
both quasiparticle bands.  

Another example that is relevant to the case of a 
2D electron gas occurs when the translational symmetry is broken
in the Wigner-crystal ground state.  The broken symmetry 
requires a modification of our analysis that we will not
detail here, since the Greens function in no longer diagonal in wavevector.
Nevertheless, the connection between
strong correlations and peaks in the spectral function
that are away from the chemical potential is again present. 
The chemical potential in this system is the energy 
increase upon adding a particle by slightly decreasing 
the lattice constant, or the energy decrease upon removing
a particle by slightly increasing the lattice constant.
The particle portion of the spectral function 
has a peak at energies far above the chemical potential
since the particle can be added at an arbitrary position, including 
energetically unfavorable positions near one of the crystal lattice sites.
The hole portion of the spectral-weight function is peaked at energies
well below the chemical potential, since the state with
a particle removed from a lattice site will have an energy substantially  
larger than that of the adjusted lattice constant state.  The correlation
energy of the Wigner crystal state is proportional to
the separation of these two non-quasiparticle 
peaks in the spectral weight function.

The 2D electron system is believed to have
a Fermi-liquid ground state at all values of $r_s$ below 
those at which the transition to the Wigner-crystal
state occurs.  For large values of $r_s$, where 
a relatively small fraction of the spectral weight 
lies in the quasiparticle peak, the 
correlation energy is very close to that of the 
Wigner-crystal state.  The electronic state is 
nearly as strongly correlated as the quantum-fluid states that occur
at strong magnetic fields in partially filled 
Landau levels.  Based on the preceeding 
examples and the sum rules discussed above, this
suggests that the non-quasiparticle portion 
of the spectral function at zero field and low
electron densities will be qualitatively 
similar to those at strong fields, leading to 
2D-2D tunneling spectroscopy results which have  
a small Fermi-liquid peak at the chemical potential
but are otherwise similar to those illustrated 
in Fig.[1].  Since $n_{\vec k} \ll 1$ for all
$\vec k$ in this density regime, it follows from 
Eq.(~\ref{correnergy}) that the non-Fermi-liquid
peak in the tunneling conductance should occur when
$eV$ is comparable to the correlation energy per particle.  
The situation should be quite close to that 
expected in Hubbard model systems where strongly dressed 
quasiparticle bands are separate from the lower and 
upper Hubbard bands.  Tunneling spectroscopy studies of 
2D electron gas systems have the potential to provide even 
more detailed information on the spectral weight function than 
what has been extracted from exhaustive angle-resolved-photoemission
studies in the cuprate superconductor case.  

\section{Summary} 

Using the spectral-weight function of the one-particle
Greens function as an example, we have attempted to 
illustrate the rich many-body physics of the 
two-dimensional electron gas system.  Compared to
other strongly correlated fermion systems, the 
2DEG has the advantage that it is accurately described
by a simple translationally invariant 
model of fermions interacting via long-range forces.
The band structure of the host semiconductor enters only
through an effective mass parameter which is 
accurately known.  At strong magnetic fields,
the 2DEG exhibits a rich variety of strong-correlation
physics that has been widely studied in connection with the 
fractional quantum Hall effect.  In this regime, 2D
tunneling-spectroscopy experiments have demonstrated that the 
spectral weight has a wide and deep gap
around the chemical potential.  The size of this gap can 
be related to the ground-state correlation energy of 
the system by invoking a first-moment sum rule for the 
spectral function.  To date, most experiments on 2D electron
systems in the absence of a field have been performed
in situations where the electrons are moderately correlated.
Experiments show clear Fermi liquid behavior, which is 
evidenced in 2D tunneling-spectroscopy experiments by a 
{\it peak} in the tunneling conductance at the Fermi energy.
The quasiparticle lifetime can be extracted from these 
measurements and is in qualitative agreement with a 
simple random-phase approximation for the electronic 
self-energy in the Fermi-liquid state.  Systems with lower 
electron densities, which are becoming available because of 
continuous advances in the epitaxial growth of 
semiconductor quantum wells, are expected to be 
more strongly correlated.  We propose that, in this regime,
the spectral function will develop peaks away from the 
Fermi energy that are analogous to both those which 
occur at strong magnetic fields and to the lower and 
upper Hubbard-band peaks thta occur in strongly correlated
lattice models.  As new frontiers in materials preparation are 
reached, it will be possible to confidently 
compare theory and experiment for extremely strongly
correlated Fermi-liquid states.  For the 2D electron gas, tunneling
spectroscopy provides a uniquely powerful probe of a strongly
correlated Fermi liquid.  The open many-body
physics questions in the low-density 2D electron gas have a large overlap 
with open issues in the many-body physics of other 
systems, and especially with the physics of planar
high-temperature-superconductor materials.\cite{elbio}

\section*{Acknowledgments}

Instructive interactions with Jim Eisenstein, Rudolf Haussmann,
Tomas Jungwirth, Hiro Mori, Sheena Murphy, and Lian Zheng are gratefully
acknowledged.  I am grateful to Ulrich Zuelicke for helpful comments 
on a draft of this paper.  This work was supported by
the National Science Foundation under grant DMR-9714055.

\eject

\end{document}